\documentclass{webofc}

 \usepackage{mathrsfs, amsmath, amssymb, amsthm, amsxtra, bm}
 \usepackage[varg]{txfonts}   

\usepackage{commands}
\usepackage{graphicx}
\usepackage{multirow}
\usepackage{csvsimple}
\usepackage{pgfplotstable, booktabs}
\usepackage{multicol}
\usepackage{todonotes}
\usepackage{xcolor}
\usepackage[caption=false]{subfig}
\usepackage{siunitx}

  \renewcommand{\d}{\mathrm{d}}

\begin{document}
\title{Power spectrum estimation methods on intracluster medium surface brightness fluctuations}
%
%

\author{\lastname{M. Bishop}\inst{1}\fnsep\thanks{e-mail: mark.bishop@vuw.ac.nz} \and
        \lastname{Y. Perrott}\inst{1} \and
        \lastname{T. Parashar}\inst{1} \and
        \lastname{S. Oughton}\inst{2}
}

\institute{Victoria University of Wellington, New Zealand
\and
           University of Waikato, New Zealand
}

\abstract{%

Accurate estimation of galaxy cluster masses is a central problem in cosmology. Turbulence is believed to introduce significant deviations from the hydrostatic mass estimates. Estimation of turbulence properties is complicated by projection of the 3D cluster onto the 2D plane of the sky, and is commonly done in the form of indirect probes from fluctuations in the X-ray surface brightness and Sunyaev-Zeldovich effect maps. In this paper, we address this problem using simulations. We examine different methods for estimating the power spectrum on 2D projected fluctuation data, emulating data projected onto a 2D plane of the sky, and comparing them to the original, expected 3D power spectrum. Noise can contaminate the power spectrum of ICM observations, so we also briefly compare a few methods of reducing noise in the images for better spectral estimation.



}
\maketitle
\section{Introduction}
\label{intro}
The total mass of a galaxy cluster is one of its most fundamental properties. 
Estimations of this mass often rely on the assumption of hydrostatic equilibrium (HSE), an approach with shortcomings since the intracluster medium (ICM) is continuously disturbed by mergers, feedback processes, and motions of galaxies. These processes generate gas motions that contribute nonthermal pressure, 
typically associated with turbulence, that leads to an underestimation of the mass by as much as 30\% \cite{simionescu_constraining_2019}. We can measure turbulence through indirect probes that come in the form of fluctuations in the X-ray surface brightness and Sunyaev-Zeldovich (SZ) effect maps. ICM observations via the X-ray and SZ effect give projected (along the line-of-sight $\ell$, onto the 2D sky-plane vector coordinate $\vec{\theta}$) 
versions of the density and pressure, respectively:
\begin{align*}
    I_{X}(\vec{\theta}) \, \propto \int n_e(\vec{\theta}, \ell)^2 \Lambda(T) \, \d\ell \; ; \quad Y_{SZ}(\vec{\theta}) \, \propto \int P_e(\vec{\theta}, \ell) \, \d\ell
 ,
\end{align*}
with some temperature dependent cooling function $\Lambda(T)$ for the X-ray emissivity \cite{sutherland_cooling_1993}. The fluctuations can be obtained using Reynolds decomposition, \ie{} $A = A_0 + \delta A$ where $A_0 = \avg{A}$ is a large scale (mean) estimate and $\delta A$ is the turbulent fluctuations.
Typically, $I_{X,0} = \avg{I_{X}}$ and $Y_{SZ,0} = \avg{Y_{SZ}}$ are represented by a circular or elliptical model \cite{churazov_x-ray_2012, khatri_thermal_2016, romero2023inferences}. Subtracting the mean profile from the observed image 
provides  the surface brightness fluctuations ($\delta I_{X}, \, \delta Y_{SZ}$), which have zero mean. These fluctuations are interpreted as 
turbulence-induced departures from the large-scale HSE model.

It is common to analyze turbulence fluctuations using a power spectrum, 
i.e., by decomposing the energy into contributions at each scale.
There are many different techniques for accomplishing this and
here we implement and compare some of those 
employed in the intracluster/interstellar medium, solar wind, and general turbulence literature. The first section will briefly discuss the different spectrum estimation methods and compare their results on projected (simulated and synthetic) fluctuation data. 
As observations contain noise contributions that interfere with power spectrum estimates, 
 we also look at different techniques for reducing the noise contribution 
 so as to
 obtain improved spectral estimates. This proceeding serves as an initial exploration into these issues. A more thorough examination and description will be published at a later date.

\section{Power Spectra: Comparison of methods}
\label{sec-1}

\begin{figure}[!hbt]%
    \centering
    \includegraphics[scale=1]{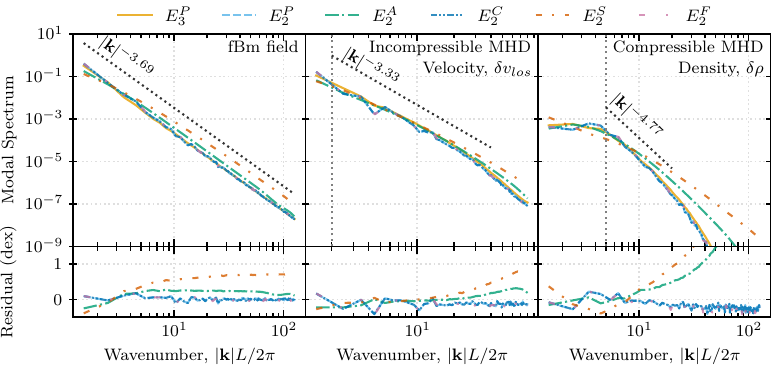}
    \caption{Comparison of the 3D (periodogram method) power spectrum, 
    $E_3^P$, to 2D projected power spectra, $E_2^X$, obtained using the periodogram, \arevalo{}, correlogram, structure function, and flatsky methods, for three different datasets.
    Left:  Using a fractional Brownian motion field, which could represent any scalar variable. Middle: The $z$-axis (line-of-sight) component of the velocity fluctuations of an incompressible MHD simulation. Right: The density fluctuations of a compressible MHD simulation. Note that the periodogram, correlogram, and flatsky methods are equivalent.}%
    \label{1:fig:sim_all_methods}%
\end{figure}

We use synthetic data including fields generated using fractional Brownian motion (fBm) and data from 3D (incompressible 
and compressible) MHD simulations. Fractional Brownian motion fields are \textit{not} turbulent but have a similar statistical structure to turbulent fields. The fBm field is generated with a 3D Hurst parameter of $H_3\approx1/2$, which becomes $H_2\approx5/6$ when projected to 2D and corresponds to a power spectrum power law of $\abs{\vec{k}}^{-11/3}$ \citep{barnsley1988science, stutzki_fractal_1998}. The astrophysical observations can provide information about density \cite{churazov_x-ray_2012}, pressure \cite{khatri_thermal_2016, romero2023inferences}, and velocity \cite{hitomi_collaboration_quiescent_2016, hitomi_collaboration_atmospheric_2018, li_direct_2020}. The synthetic data that we use represent these variables. The fBm field could represent any of these.
For the incompressible MHD simulation, we use the line-of-sight component of the velocity ($v_{los}$).

For each synthetic dataset we calculate the 3D modal spectrum \cite{oughton2015anisotropy} averaged over spherical surfaces at radius $\abs{\vec{k}}$ (in dimensionless units relative to the box size $L$, which could be in units of \si{Mpc} or \si{kpc} for a real observation), $E_{3}^{P}(\abs{\vec{k}})$. This is commonly called the power spectrum in astronomical literature \cite{brunt_intrinsic_2003}. These are used as the `ground truth' to compare against various 2D spectra of the fluctuations projected along the line-of-sight. The 2D modal spectra are averaged over circular annuli of radius $\abs{\vec{k_\theta}}$. The five spectral estimation methods we test are:



\textit{Periodogram:} The periodogram spectral estimate is simply the squared magnitude of the Fourier transform of the data \cite{stoica_spectral_2005}, denoted with the superscript $P$, $E^{P}$. 

\textit{Correlogram:} The correlogram spectral estimator ($E^{C}$) takes the Fourier transform of the auto-correlation function (called Blackman-Tukey if a window function is applied) \cite{blackman_measurement_1960, stoica_spectral_2005}. The Weiner-Khinchin theorem states that this and the periodogram give the same result.

\textit{\arevalo{}:} A scale-space decomposition via Gaussian convolutions, with the standard deviation of the Gaussian approximating the scale, can be used to estimate the power spectrum ($E^{A}$) \cite{arevalo_mexican_2012}. This is the most common method used in the ICM turbulence literature \cite{churazov_x-ray_2012, zhuravleva_suppressed_2019, romero2023inferences}.

\textit{Structure function:} A structure function \cite{davidson_turbulence_2005} describes the moments of a lagged difference and loosely represents the variance as a function of a length scale (the lag $\ell$). A crude approximation can turn the second order structure function into an equivalent power spectrum ($E^{S}$) as a function of 
equivalent wavenumber ($k \equiv 2 \pi/\ell$)  \cite{davidson_turbulence_2005, chhiber_higher-order_2018}.

\textit{Spherical harmonics:} Given that the ICM is observed on the sky-sphere, a spectral decomposition using spherical harmonics is natural. The small size of the cluster allows for a flat-sky approximation \cite{khatri_thermal_2016} to estimate the power spectrum ($E^{F}$).

\pgfkeys{/pgf/number format/.cd, sci, sci zerofill, precision=2}
\newcommand{\STAB}[1]{\begin{tabular}{@{}c@{}}#1\end{tabular}}
\begin{table}
\centering
\begin{tabular}{|c}
    \toprule
    \\
    \midrule
    \parbox[t]{1mm}{\multirow{3}{2mm}{\rotatebox[origin=c]{90}{fBm}}}\\
    \\
    \\
    \midrule
    \parbox[t]{1mm}{\multirow{3}{2mm}{\rotatebox[origin=c]{90}{IMHD}}}\\
    \\
    \\
    \midrule
    \parbox[t]{1mm}{\multirow{3}{2mm}{\rotatebox[origin=c]{90}{CMHD}}}\\
    \\
    \\
    \bottomrule
\end{tabular}
\begin{filecontents*}{data/method_stats.csv}
sim,Method,Methodsym,MSE,MSEdex,MAPE,MAPEdex
fBm,Periodogram,$E^{P}_{2}(\abs{\vec{k_\theta}})$,5.32e-05,0.00135,0.0592,0.00695
fBm,Arevalo,$E^{A}_{2}(\abs{\vec{k_\theta}})$,0.000159,0.0396,0.567,0.0384
fBm,SF,$E^{S}_{2}(\abs{\vec{k_\theta}})$,0.00029,0.452,3.71,0.116
IMHD,Periodogram,$E^{P}_{2}(\abs{\vec{k_\theta}})$,3.93e-05,0.0197,0.249,0.0291
IMHD,Arevalo,$E^{A}_{2}(\abs{\vec{k_\theta}})$,2.75e-05,0.0411,0.517,0.0333
IMHD,SF,$E^{S}_{2}(\abs{\vec{k_\theta}})$,3.92e-05,0.251,2.17,0.0766
CMHD,Periodogram,$E^{P}_{2}(\abs{\vec{k_\theta}})$,1.35e-09,0.0556,0.39,0.0227
CMHD,Arevalo,$E^{A}_{2}(\abs{\vec{k_\theta}})$,2.66e-10,5.39,1380.0,0.173
CMHD,SF,$E^{S}_{2}(\abs{\vec{k_\theta}})$,4.62e-09,12.2,62100.0,0.27
\end{filecontents*}
\pgfplotstabletypeset[
    col sep=comma,
    brackets/.style={
        postproc cell content/.append style={/pgfplots/table/@cell content/.add={\relax(}{)}},
    },
    columns={Method, MSE, MSEdex, MAPE, MAPEdex},
    columns/sim/.style={column name=\,, string type, column type/.add={}{|}},
    columns/Method/.style={string type, column type/.add={}{||}},
    columns/MSE/.style={column type={r}},
    columns/MSEdex/.style={column name=(dex), brackets, column type={l}},
    columns/MAPE/.style={column type={r}},
    columns/MAPEdex/.style={column name=(dex), brackets, column type={l|}},
    every first column/.style={column type/.add={|}{}},
    every last column/.style={column type/.add={}{|}},
    every head row/.style={before row=\toprule, after row=\midrule},
    every last row/.style={after row=\bottomrule},
    every row no 3/.style={before row=\midrule},
    every row no 6/.style={before row=\midrule},
]{data/method_stats.csv}
\caption{The MSE and MAPE statistical measures of the accuracy of the 2D power spectra versus the true 3D power spectrum for the different spectrum estimation methods shown in figure~\ref{1:fig:sim_all_methods}: a fBm field, an incompressible MHD simulation (IMHD), and a compressible MHD simulation (CMHD).}
\label{1:tab:ssim_all_methods}
\end{table}



Figure~\ref{1:fig:sim_all_methods} shows the true, averaged modal spectrum for the fluctuations, $E_{3}^{p}$, and the above five 2D proxies for $E_{3}^{P}$. These are calculated for: (left) a 3D fBm field with a spectral slope of $\abs{\vec{k_3}}^{-11/3}$, where $\abs{\vec{k_3}}$ is the magnitude of the 3D wavevector, (middle) an incompressible MHD simulation, and (right) a compressible MHD simulation. The plots at the bottom of the figure~\ref{1:fig:sim_all_methods} show the residual (difference) in units of order of magnitude (dex) of the 2D projected power spectra (scaled by a factor of $\Delta k/2 \pi$ to account for the difference in dimensions) compared to the true 3D spectrum.

The projection-slice theorem states that the Fourier transform of the 2D projected data should be a slice in the Fourier transform of the original 3D data (\ie{}the line-of-sight wavenumber $k_\ell = 0$). Under the assumption of isotropy, this implies same modal spectra for 3D and 2D projected. The Fourier methods---periodogram, correlogram, and flatsky---yield
the same results:
 their 2D and 3D spectra are equivalent. Evidently, this is not the case for the structure function and \arevalo{} spectral estimates. These two methods obtain an appropriate powerlaw for the fBm field, and for the inertial range in the incompressible and compressible MHD simulations (indicated by the black dotted lines in Fig.~\ref{1:fig:sim_all_methods}). The \arevalo{} spectral estimate has a known bias for a pure powerlaw spectrum that can be corrected for, if the powerlaw of the 3D 
 spectrum is already known \cite{arevalo_mexican_2012, romero2023inferences}. This bias does not necessarily account for the order of magnitude difference seen in the compressible MHD case beyond the inertial range. This has interesting implications for the capabilities of the \arevalo{} method to actually capture the dissipation scale \cite{zhuravleva_suppressed_2019}. Similarly, the structure function method can obtain reasonably correct powerlaws, but has significant bias in the fBm field, and compressible MHD beyond the inertial range.

To further quantify the accuracy of these methods, we compute the mean squared error (MSE) and mean absolute percentage error (MAPE) for each 2D power spectrum ($E^{X}_{2}(\abs{\vec{k_{\theta}}})$). These are listed in table~\ref{1:tab:ssim_all_methods}.
\begin{align*}
    \text{MSE} = \avg{\bracket{E^{X}_{2}(\abs{\vec{k_{\theta}}}) - E^{P}_{3}(\abs{\vec{k_3}})}^2} ; \quad
    \text{MAPE} = \avg{\abs{\frac{E^{X}_{2}(\abs{\vec{k_{\theta}}}) - E^{P}_{3}(\abs{\vec{k_3}})}{E^{P}_{3}(\abs{\vec{k_3}})}}}
\end{align*}
where the angle brackets $\avg{\cdot}$, represent the average over available wavenumbers. We also calculate the MSE and MAPE in units of \textit{dex}, which has the same formula except we take the base-10 log of $E^{X}_{2}(\abs{\vec{k_{\theta}}})$ and $E^{P}_{3}(\abs{\vec{k_3}})$. 
The power spectrum values decrease by orders of magnitude from small to large wavenumbers. This biases the MSE on large-scale accuracy. MAPE describes the relative error, and is therefore scale-independent. By using units of \textit{dex}, we look at the order of magnitude differences, which can rebalance the error estimates, particularly for the MSE. The MSE for IMHD and CMHD in table~\ref{1:tab:ssim_all_methods} indicates that the \arevalo{} method is the best, which is counter to what figure~\ref{1:fig:sim_all_methods} shows. However MSE (dex) agrees with both the MAPE and MAPE (dex) orderings, \arevalo{} is not the best for these cases. As indicated 
by figure~\ref{1:fig:sim_all_methods}, periodogram works the best, followed by \arevalo{}. The structure function equivalent spectrum has the largest error, especially 
for wavenumbers 
in the dissipation range of the MHD simulations.




\section{Noise reduction methods}

\begin{figure}[!hbt]%
    \centering
    \includegraphics[scale=1]{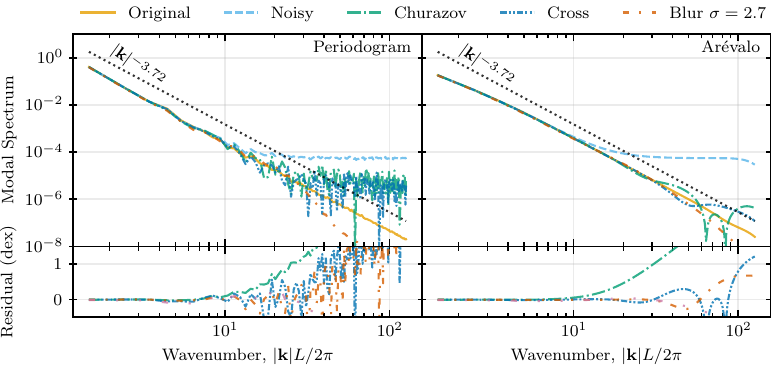}
    \caption{Comparison of the original (noiseless) fBm spectrum to the version with added Poisson noise, and the versions obtained using three different noise removal methods. 
    Spectra are calculated
    using the periodogram (left) and \arevalo{} (right) methods. 
    }%
    \label{1:fig:fbm_noise_methods}%
\end{figure}

We now address the issue of Poisson noise, which is typically introduced by the counting of photons by the telescope detectors.
We \textit{apply} Poisson noise to a (noiseless) fluctuation image by treating each pixel as the mean (and therefore, the variance) of the Poisson distribution to sample from. The noise reduction methods described below can be applied to any method described in section~\ref{sec-1}; 
here we only show results for the periodogram and \arevalo{} methods. We test the following noise reduction methods:

\textit{Churazov et al:} From a noisy observation (with a single realization of the Poisson noise), we can generate additional realizations by sampling each (noisy) pixel as the mean and variance of the Poisson distribution. This method requires generating many (we use 100) additional realizations of the noisy image and averaging their power spectrum. The difference between the original noisy spectrum and the realization-averaged power spectrum gives an estimate of the (constant) noise floor. Subtracting the noise floor from the original noisy spectrum obtains an estimate on the noiseless spectrum \cite{churazov_x-ray_2012}. 

\textit{Cross spectrum:} Data are split into two observations $A$ and $B$; either by independent observations, or splitting one observation cube into two parts, and the cross-power spectrum is computed. It retains the spectral properties of features common to the two observations whereas the \textit{independent} noise realizations cancel out. This is equivalent to the `jack-knife' method: $(A+B)/2$ contains the data of interest and $(A-B)/2$ contains the noise component \cite{cerini2023new}. Cross spectra can be computed for all the methods in section~\ref{sec-1} (see \citep{romero2023inferences} for an \arevalo{} cross-spectrum method).

\textit{Gaussian blur:} Lastly, a simple Gaussian blur with some selected scale, ideally one at which the signal to noise ratio (SNR) drops to a predetermined value, can be used to remove noise. This blurs the fluctuations below the selected scale, and retains structures at larger scales. However, the choice of an appropriate blurring scale (Gaussian $\sigma$) is difficult without knowledge of the true spectrum. For the case shown here, we iterate over different scales and select the value that 
best reconstructs the original spectrum. 

Amongst these methods, the cross-spectrum and blurring methods are the simplest to implement and understand. We apply all methods to the fractional Brownian motion fields with the goal of recreating the original 2D (projected) spectrum. We generate two realizations of the noise onto the original image for use with the cross-spectrum. The results are shown in figure~\ref{1:fig:fbm_noise_methods} and table~\ref{1:tab:fbm_noise_methods}, where we show the periodogram and \arevalo{} methods of power spectrum estimation (left and right respectively).

For both the periodogram and \arevalo{} methods, the noisy spectra are 
essentially equal
 to
the original spectrum, until the spectral level drops to that of the constant noise floor, as is expected. All noise reduction methods tested  are able to recover more of the original spectral range compared than is available in the noisy spectrum. The extent/range recovered depends on the amount of noise and the method used. Table~\ref{1:tab:fbm_noise_methods} shows that the cross spectrum and Gaussian smoothing techniques seem to work best in spite of their simplicity. For the \arevalo{}-produced spectra, both these methods yield comparable results, while for periodogram-produced spectra Gaussian smoothing produces a smoother end result. 




\begin{table}
\centering
\begin{tabular}{|c}
    \toprule
    \\
    \midrule
    \parbox[t]{1mm}{\multirow{4}{1mm}{\rotatebox[origin=c]{90}{P.gram}}}\\
    \\
    \\
    \\
    \midrule
    \parbox[t]{1mm}{\multirow{4}{1mm}{\rotatebox[origin=c]{90}{Ar\'{e}valo}}}\\
    \\
    \\
    \\
    \bottomrule
\end{tabular}
\begin{filecontents*}{data/all_noise_stats.csv}
SpecType,Method,MSE,MSEdex,MAPE,MAPEdex
Periodogram,Noisy,6.27e-08,5.47,608.0,0.307
Periodogram,Churazov,9.63e-09,2.2,64.4,0.18
Periodogram,Cross,6.07e-08,1.74,38.4,0.157
Periodogram,Smooth,9.87e-07,0.962,0.736,0.126
Arevalo,Noisy,4.03e-08,4.62,350.0,0.285
Arevalo,Churazov,2.1e-08,0.3,2.23,0.0533
Arevalo,Cross,3.9e-08,0.183,1.54,0.0475
Arevalo,Smooth,3.01e-07,0.262,0.485,0.0576
\end{filecontents*}
\pgfplotstabletypeset[
    col sep=comma,
    brackets/.style={
        postproc cell content/.append style={/pgfplots/table/@cell content/.add={\relax(}{)}},
    },
    columns={Method, MSE, MSEdex, MAPE, MAPEdex},
    columns/Method/.style={string type, column type/.add={}{||}},
    columns/MSE/.style={column type={r}},
    columns/MSEdex/.style={column name=(dex), brackets, column type={l}},
    columns/MAPE/.style={column type={r}},
    columns/MAPEdex/.style={column name=(dex), brackets, column type={l|}},
    every first column/.style={column type/.add={|}{}},
    every last column/.style={column type/.add={}{|}},
    every head row/.style={before row=\toprule, after row=\midrule},
    every last row/.style={after row=\bottomrule},
    every row no 4/.style={before row=\midrule},
]{data/all_noise_stats.csv}
\caption{The MSE and MAPE statistical measures of the 2D projected noiseless power spectra versus noise reduction methods for the fBm field shown in figure~\ref{1:fig:fbm_noise_methods}.}
\label{1:tab:fbm_noise_methods}
\end{table}

\section{Conclusion}
We compare different spectrum calculation techniques, constructed in very different ways, with different pros and cons. The differences in the outputs
of these methods could potentially introduce biases in physical interpretations. There is a clear advantage to the Fourier methods (periodogram, correlogram, and flatsky), but these methods can behave poorly when data are masked, which is common for astrophysical observations. Sharp discontinuities introduced by masking create aliasing effects in the wavenumber space, polluting the spectral estimates \citep{arevalo_mexican_2012}. \arevalo{} and structure function methods are more robust to missing data because they rely on averaging in lag or scale space. However, they struggle to capture the dissipation/kinetic range of the spectrum. Simple minded noise reduction techniques appear to work reasonably well in extending the spectral range below the noise floor. Future work will expand on the information 
presented in this proceeding.


%
%
%
%
%


\end{document}